\documentclass[a4paper]{article}

\usepackage{amsthm,amsmath,amssymb,amsfonts}

\def\bbbn{{\mathbb N}}
\def\bbbc{{\mathbb C}}

\def\cp1{{\mathbb C\mathbb P}^1}
\def\cA{{\cal A}}
\def\cU{{\cal U}}

\newtheorem{Def}{Definition}

\newtheorem{Thm}{Theorem}
\newtheorem{Pro}{Proposition}

\newtheorem{Rem}{Remark}
\newtheorem{Ex}{Example}

\begin{document}
\bibliographystyle{alpha}
\title{On the structure of $(2+1)$--dimensional commutative and noncommutative
integrable equations}
\author{Jing Ping Wang\\
Institute of Mathematics and Statistics, University of Kent, UK}
\maketitle

\begin{abstract}
We develop the symbolic representation method to derive
the hierarchies of $(2+1)$-dimensional integrable equations from the  
scalar Lax operators and to study their properties globally. 
The method applies to both commutative and noncommutative 
cases in the sense that the dependent variable takes its values in $\bbbc$ 
or a noncommutative associative algebra.  
We prove that these hierarchies are indeed quasi-local in the commutative case
as conjectured by  Mikhailov and Yamilov in 1998, \cite{mr1643816}.
We propose a ring extension based on the symbolic representation. 
As examples, we give noncommutative versions of 
KP, mKP and Boussinesq equations.
\end{abstract}

\section{Introduction}\label{Sec1}
Integrable $(1+1)$--dimensional nonlinear evolution equations, i.e.
equations of the form
\[ u_t=F(u,u_x,u_{xx},...,u_{xx...x})\,  \] 
possess rich algebraic and geometrical structures such as
Lax representations, a bi-Hamiltonian formulation,
infinitely many symmetries and  conservation laws, etc.
A lot of work has been devoted to the study of such equations and
comprehensive classification results have been obtained.

The extension of  these remarkable structures  to the $(2+1)$--dimensional case
is not straightforward.
It was proved in \cite{ZK84} that there is no
bi-Hamiltonian formulation for $(2+1)$--dimensional
integrable equations of the same type as those for 
the $(1+1)$--dimensional case like the Korteweg-de Vries (KdV) equation
$$u_t=u_{xxx}+6 u u_x.$$
In 1988, Fokas and Santini, cf. \cite{FS88c,FS88b}, constructed a bi-Hamiltonian
structure for the Kadomtsev-Petviashvili (KP) equation
$$ u_t=u_{xxx}+6 u u_x +3 D_x^{-1} u_{yy} $$
by considering it as a reduction of $(3+1)$-dimensional system.
Meanwhile, Magri and his coauthors, \cite{MMT88}, explained this structure
from the geometric point of the view by developing the concept of
Nijenhuis G-manifolds,
which amazingly works for both one and two space dimensions.
Dorfman and her coauthors 
introduced the noncommutative ring of formal pseudo-differential
operators, cf. \cite{mr93c:58085,mr94h:35241}.
They proved that the Fokas-Santini bi-Hamiltonian structure of the KP
can be obtained from the Adler-Gel'fand-Dikii (AGD) scheme
by considering the second order Lax operator
$$
L=D_x^2+u-\partial_y.
$$
The bi-Hamiltonian structure naturally leads to a recursion operator for
the equation.
However, the operator (so-called operand) takes its value in the ring of
formal differential operators and is of a different type from
the one for KdV (and therefore does not contradict the result of \cite{ZK84}).
A hierarchy of infinitely many symmetries
can only be produced by making suitable combinations of
the operator acting on distinct seeds \cite{MMT88}. 

One of the main obstacles to extend the spectacular results 
of $(1+1)$-dimensional integrable equations to the $(2+1)$-dimensional
case is that the equations themselves,
their higher symmetries and conservation laws are non-local, i.e.
the appearance of the formal integral $D_x^{-1}$ or $D_y^{-1}$. 
In 1998, Mikhailov and Yamilov, \cite{mr1643816}, introduced the concept
of quasi-local functions based on the observation that
the operators $D_x^{-1}$ and $D_y^{-1}$ never appear alone
but always in pairs like $D_x^{-1}D_y$ and $D_y^{-1}D_x$ for
all known integrable equations and their hierarchies of symmetries 
and conservation laws, which enables them to extend the symmetry approach
of testing integrability \cite{mr93b:58070} to the $(2+1)$-dimensional case. 

In this paper, we develop the symbolic representation method to produce
hierarchies of $(2+1)$-dimensional integrable equations from
scalar Lax operators and to study their properties globally. 
The method applies both to the commutative and noncommutative 
case in the sense that the dependent variable takes its values in $\bbbc$ 
or in a noncommutative associative algebra.  
We prove that these hierarchies are indeed quasi-local in the commutative case
as conjectured by  Mikhailov and Yamilov in 1998, \cite{mr1643816}.
This concept of quasi-locality has to be extended in the noncommutative case. 
Here we propose a ring extension based on the symbolic representation.
As examples, we give noncommutative versions of 
the KP, mKP and Boussinesq equations.
\section{Quasi-local polynomials and symbolic representation}\label{Sec2}
The symbolic method was introduced by Gel'fand and Dikii in 
1975 \cite{mr58:22746}.
However, its basic idea can be dated back to the middle of 19th
century. Recently, we successfully applied this method to 
the classification of both commutative and noncommutative
$(1+1)$-dimensional homogeneous evolution equations in a 
series of papers, cf. \cite{mr99g:35058,mr99i:35005,mr1781148}.
With the help of number theory, it enables us to give 
a global description of their integrable hierarchies \cite{wang98}.
The symbolic method is also powerful in dealing with differential
(cf. \cite{mr1923781}) and pseudo-differential (cf. \cite{mr1908645}) 
operators. 
The authors of  \cite{mr1908645} generalised the standard symmetry
approach  \cite{mr93b:58070} and made it suitable for the
study of nonlocal and non--evolutionary equations, see also \cite{mnw06}.

In this section, we'll extend the symbolic method to the case of
two spatial variables $x$ and $y$. For simplicity, we restrict our
attention to a single dependent variable $u$.
Extensions to several independent variables and dependent variables 
are straightforward.
\subsection{Quasi-local polynomials}
We begin with basic definitions and notations of the ring of commutative and 
non-commutative differential polynomials. 
The derivatives of dependent variable $u$ with respect to its
independent variables $x$ and $y$ are denoted by $u_{ij}=\partial_x^i 
\partial_y^j u$. For smaller $i$ and $j$, we sometimes write the indices out explicitly,
that is $u_{xxy}$ and $u$ instead of $u_{21}$ and $u_{00}$.
A differential monomial takes the form
$$u_{i_1 j_1}u_{i_2 j_2}\cdots u_{i_k j_k} \  .$$
We call $k$ the degree of the monomial.
We let $\cU^k$ denote the set of differential polynomials of degree $k$.
The ring of differential polynomials is denoted by 
$\cU=\oplus_{k\ge 1}\cU^k$. Notice that $1\notin \cU$ and
it is a differential ring with total $x$-derivation and $y$-derivation
$$
D_x=\sum_{i=0}^{+\infty}\sum_{j=0}^{+\infty}u_{i+1,j}\frac{\partial}
{\partial u_{ij}}
\quad \mbox{and}\quad 
D_y=\sum_{i=0}^{+\infty}\sum_{j=0}^{+\infty}u_{i,j+1}\frac{\partial}
{\partial u_{ij}}.
$$
The ring $\cU$ is commutative if the dependent variable $u$ 
takes its values in a commutative algebra, for example the ring
of smooth functions in \(x\) and \(y\).
Let us denote
\begin{eqnarray}\label{Theta}
\Theta=D_x^{-1}D_y, \qquad \Theta^{-1}=D_y^{-1}D_x.
\end{eqnarray}
The concept of {\bf quasi-local (commutative) polynomials $\cU(\Theta)$} 
was introduced in \cite{mr1643816} to test integrability of
a given equation.
To define it, we consider a sequence of extensions of $\cU$. 
Let $\Theta \cU=\{\Theta f:\
f \in \cU\}$ and $\Theta^{-1} \cU=
\{\Theta^{-1} f:\ f \in \cU\}$. We define
$\cU_0(\Theta )=\cU$
and $\cU_k (\Theta)$ is the ring closure of the union 
$$\cU_{k-1}(\Theta )\cup \Theta \cU_{k-1}(\Theta ) 
\cup \Theta^{-1} \cU_{k-1}(\Theta ).$$
Here the index $k$ indicates the maximal depth
of nesting the operator $\Theta^{\pm1}$ in the expression.
Clearly, we have $\cU_{k-1}(\Theta )\subset \cU_k(\Theta )$.
We now define $\cU(\Theta)=\lim_{k\to \infty}\cU_k (\Theta)$. 
However, for a given $f\in \cU(\Theta)$, there exists $k$
such that $f\in \cU_k(\Theta)$.
The extension $\cU(\Theta)$ has a natural gradation according to the 
number of $u$ and its derivatives
$$\cU(\Theta)=\oplus_{l\ge 1}\cU^l(\Theta),\quad 
\cU^p(\Theta)\cdot\cU^q(\Theta)\in\cU^{p+q}(\Theta).$$

Note that $\cU_k(\Theta)$ is not invariant under transformation
of variables. For example, a simple transformation
\begin{eqnarray}\label{TR}
x'=x+y, \qquad y'=y
\end{eqnarray}
transforms $\Theta=D_{x'}^{-1} D_{y'} \mapsto 1-\Theta=1-D_x^{-1} D_y$. 
Hence $\Theta^{-1} \mapsto (1-\Theta)^{-1}$, which is not in $\cU_k(\Theta)$.

If $u$ takes its value in a noncommutative associative algebra, the ring $\cU$ 
is noncommutative. Typical examples
are the algebras of $n\times n$ matrices and Clifford algebras
(see \cite{mr99c:58077} for more examples and noncommutative $(1+1)$-dimensional
integrable evolution equations). For any element $f$, $g$ and $h$ in a
noncommutative associative algebra, we use the notation
\begin{eqnarray}\label{Mut}
L_f (g)=fg, \qquad R_f(g)=gf
\end{eqnarray}
for the operators of left and right multiplication and 
\begin{eqnarray}\label{cmt}
C_f (g) =L_f (g)-R_f(g) =fg-gf
\end{eqnarray}
for the commutator (notation $ad_f$ is also commonly used). 
It is a derivation since it satisfies 
the Leibniz rule: $C_f (gh)=C_f (g) h+g C_f (h)$.


So far there is no proper extension from 
noncommutative differential polynomials
to quasi-local noncommutative polynomials,
mainly because the study of 
noncommutative $(2+1)$-dimensional integrable equations is a new 
and challenging topic. 
We can take the corresponding extension of $\cU(\Theta)$ as the starting
point. However, this turns out to be too restrictive. This will
be discussed further in Section \ref{Sec42}. 
\subsection{Symbolic representation} 
The basic idea of the symbolic representation
is to replace $u_{ij}$ by $u \xi^i \eta^j$, where $\xi$ and
$\eta$ are symbols. Now the total differentiation with respect to $x$,
that is, mapping $u_{ij}$ to $u_{i+1,j}$, is replaced by multiplication with
$\xi$, as in the Fourier transform. Similarly, the total differentiation
with respect to $y$ mapping $u_{ij}$ to $u_{i,j+1}$, is replaced by 
multiplication with $\eta$. For higher degree terms with multiple $u$'s, 
one uses different symbols to denote each of them. Its 
symbolic representation depends on where $u$ takes its value,
i.e. whether it is commutative or not.
For example, the noncommutative binomial $u_{ij}u_{kl}$
has symbolic representation $u^2 \xi_1^i \eta_1^j\xi_2^k \eta_2^l$. 
In the commutative case, one has $u_{ij}u_{kl}=u_{kl}u_{ij}$.
We therefore need to average its symbolic representation over 
the permutation group $\Sigma_2$ so that $u_{ij}u_{kl}$ and $u_{kl}u_{ij}$ 
have the same symbolic form.
So the symbolic representation of $u_{ij}u_{kl}$ is
$\frac{u^2}{2} (\xi_1^i \eta_1^j\xi_2^k \eta_2^l
+\xi_2^i \eta_2^j\xi_1^k \eta_1^l)$. 
Symmetrisation makes the symbol representation of monomials unique.

Let $\cA^k$ be the space of algebraic polynomials $f$ in $2k$ variables,
$\xi_i$ and $\eta_i$, where $i=1, \cdots, k$.

The symbolic representation defines a linear isomorphism between the space
${\cU}^k$ of (non)-commutative differential polynomials of degree $k$
and the space $\cA^k$.
It is uniquely defined by its action on monomials.
\begin{Def}
The symbolic representation of a differential monomial is defined as
$$u_{i_1,j_1}u_{i_2,j_2}\cdots u_{i_k,j_k}\longmapsto 
\left\{\begin{array}{ll}
u^k \xi_1^{i_1}\eta_1^{j_1}\xi_2^{i_2}\eta_2^{j_2}\cdots 
\xi_k^{i_k}\eta_k^{j_k}, & (\mbox{noncommutative});\\
\frac{u^k}{k!} \sum_{\sigma\in \Sigma_k} \xi^{i_1}_{\sigma(1)}
\eta^{j_1}_{\sigma(1)}\cdots \xi^{i_k}_{\sigma(k)}\eta^{j_k}_{\sigma(k)},
& (\mbox{commutative}).
\end{array}\right.$$
\end{Def}
There are two parts in the symbolic representation of a monomial: the first
part, $u^k$, indicating its degree; the second part being in $\cA^k$.
The symbols of Gel'fand and Dikii only contain the second part. 
We emphasise that the first part is very important in dealing with the
monomials $u^l$ the case of several
noncommutative dependent variables.

In general, we denote the symbolic representation of $P\in \cU^k$,
whether it is commutative or not, by $\hat P$
and $Q\in \cU^l$ by $\hat Q$.
The multiplication of $P$ and $Q$ corresponds to the symbol
\begin{eqnarray}\label{nsymM}
{\widehat {PQ}}(\xi_1,\eta_1,\cdots ,\xi_k, \eta_k, \xi_{k+1},\eta_{k+1},
\cdots, \xi_{k+l},\eta_{k+l})
=\hat P(\xi_1,\eta_1,\cdots ,\xi_k, \eta_k)
\hat Q(\xi_{k+1},\eta_{k+1}, \cdots, \xi_{k+l},\eta_{k+l});
\end{eqnarray}
when $P$ and $Q$ are commutative differential polynomials, 
the right-hand side needs to be symmetrised
\begin{eqnarray}\label{symM}
&&{\widehat {PQ}}(\xi_1,\eta_1,\cdots ,\xi_k, \eta_k, \xi_{k+1},\eta_{k+1},
\cdots, \xi_{k+l},\eta_{k+l})
\nonumber\\&=&
\frac{1}{(k+l)!} \sum_{\sigma\in \Sigma_{k+l}} 
\hat P(\xi_{\sigma(1)},\eta_{\sigma(1)},\cdots ,\xi_{\sigma(k)}, 
\eta_{\sigma(k)})
\hat Q(\xi_{{\sigma(k+1)}},\eta_{{\sigma(k+1)}}, \cdots, 
\xi_{{\sigma(k+l)}},\eta_{{\sigma(k+l)}}).
\end{eqnarray}

The total derivatives $D_x$ and $D_y$ have the following representations:
\begin{eqnarray*}
&&{\widehat {D_x P}}(\xi_1,\eta_1, \xi_2, \eta_2,\cdots, \xi_k,\eta_k)
=(\xi_1+\cdots +\xi_k)
 \hat P (\xi_1,\eta_1, \xi_2, \eta_2,\cdots, \xi_k,\eta_k);
\\&&
{\widehat {D_y P}}(\xi_1,\eta_1, \xi_2, \eta_2,\cdots, \xi_k,\eta_k)
=(\eta_1+\cdots +\eta_k)
 \hat P (\xi_1,\eta_1, \xi_2, \eta_2,\cdots, \xi_k,\eta_k).
\end{eqnarray*}

Naturally, the actions of $\Theta=D_x^{-1}D_y$ and $\Theta^{-1}$
on $P\in \cU^k$ can be represented as
\begin{eqnarray*}
&&{\widehat {\Theta P}}(\xi_1,\eta_1, \xi_2, \eta_2,\cdots, \xi_k,\eta_k)
=\frac{\xi_1+\cdots +\xi_k}{\eta_1+\cdots +\eta_k}
 \hat P (\xi_1,\eta_1, \xi_2, \eta_2,\cdots, \xi_k,\eta_k);
\\
&&{\widehat {\Theta^{-1} P}}(\xi_1,\eta_1, \xi_2, \eta_2,\cdots, \xi_k,\eta_k)
=\frac{\eta_1+\cdots +\eta_k}{\xi_1+\cdots +\xi_k}
 \hat P (\xi_1,\eta_1, \xi_2, \eta_2,\cdots, \xi_k,\eta_k).
\end{eqnarray*}
Similar to the multiplication of $P$ and $Q$, we have
\begin{Def}
Let $P\in \cU^k$ and $Q\in \cU^l$. Then
\begin{eqnarray*}
{\widehat {P\Theta Q}}&=&\hat P(\xi_1,\eta_1,\cdots, \xi_k,\eta_k)
\frac{\xi_{k+1}+\cdots +\xi_{k+l}}{\eta_{k+1}+\cdots +\eta_{k+l}}
 \hat Q (\xi_{k+1},\eta_{k+1},\cdots, \xi_{k+l},\eta_{k+l}).
\end{eqnarray*}
When $P$ and $Q$ are commutative differential polynomials, 
the right-hand side needs to be symmetrised.
\end{Def}
In the same way, we can uniquely define the symbol of $P\Theta^{-1} Q$.
Together with the multiplication rule (\ref{nsymM}) or (\ref{symM}), 
we have now completely determined the symbolic representations of the elements
in $\cU_1 (\Theta)$.
By induction, we can define the symbolic representation
of any element in $\cU(\Theta)$,
which is a rational function with
its denominator being the products of the linear factors. 
The expression of the denominator uniquely determines how the
operator $\Theta^{\pm1}$ is nested. 
\begin{Ex} The expressions $u^3 \xi_1 \frac{\eta_2+\eta_3}
{\xi_2+\xi_3}\frac{\xi_3}{\eta_3}$  and $u^3\xi_1 \frac{\eta_2}
{\xi_2}\frac{\xi_3}{\eta_3}$ are the symbolic representations of
$u_{x}\Theta (u\Theta^{-1} u)\in \cU_2(\Theta)$ and
$u_{x}(\Theta u) \Theta^{-1} u\in \cU_1(\Theta)$ respectively.
\end{Ex}
Let us define the symbolic representations of pseudo-differential operators
motivated by their Fourier transforms.
First we assign a special symbol $X$ to the operator $D_x$. 
Its formal inverse $D_x^{-1}$ has the symbol $\frac{1}{X}$.
Then we have the following rules for
$f\in \cU^k(\Theta)$ with symbol $\hat f$
and $g\in \cU^l(\Theta)$ with symbol $\hat g$:
\begin{eqnarray*}
X\circ \hat f(\xi_1,\eta_1,\ldots,\xi_k,\eta_k)=
(\xi_1+\cdots+\xi_k+X)\hat f(\xi_1,\eta_1,\ldots,\xi_k,\eta_k),
\end{eqnarray*}
which represents the Leibniz rule $D_x\circ f=D_x(f)+f\circ D_x$; and the 
composition rule
\begin{eqnarray*}
&&{\hat f} X^p\circ {\hat g} X^q=
{\hat f}(\xi_1,\eta_1,\ldots,\xi_k,\eta_k)(\xi_{k+1}+\cdots+\xi_{k+l}
+X)^p {\hat g}(\xi_{k+1},\eta_{k+1},\ldots,\xi_{k+l},\eta_{k+l})X^q.
\end{eqnarray*}
This composition rule is valid for both positive and negative powers
$ p$ and $q$. For positive powers, it is a polynomial of $X$.
For negative powers, one can expand it at $X = \infty$ 
to identify it.
\begin{Ex}
The symbol of $D_x^{-1} u$ is $\frac{u}{X+\xi}$ if we assign $\xi$
as the symbol for $u$. Then in the noncommutative case we have
$$
D_x^{-1}u D_x^{-1} u=u^2 D_x^{-2}-(2u u_x+u_x u)D_x^{-3}
+(3 u u_{xx}+3 u_x^2+u_{xx} u) D_x^{-4}+\cdots\cdots
$$
Symbolically, we can compute it by
$$
\frac{u}{X+\xi} \circ \frac{u}{X+\xi}
=\frac{u^2}{(X+\xi_1+\xi_2)(X+\xi_2)}
=u^2 (\frac{1}{X^2}-\frac{\xi_1+2\xi_2}{X^3}
+\frac{\xi_1^2+3 \xi_1 \xi_2+3 \xi_2^2}{X^4}
+\cdots ).
$$
\end{Ex}
To extend the symbolic representation from one dependent variable
to several noncommutative dependent variables is straightforward. 
We need to assign new symbols for each of them such as assigning 
$\xi^{(1)}, \eta^{(1)}$ for $u$ and $\xi^{(2)}, \eta^{(2)}$ for $v$ and so on.
As one can imagine, the abstract formulas for arbitrary $n$ dependent
variables get very long
and it is difficult to present them in a compact way.
We have no problems in handling small $n$ or  given expressions.
\begin{Ex}
The symbolic representation of $u_{x}v_{y}$ is $uv\xi^{(1)}\eta^{(2)}$
and that of $v_{y}u_{x}$ is $vu\xi^{(1)}\eta^{(2)}$.
\end{Ex}
Notice that the only difference in the symbolic representations for
$u_{x}v_{y}$ and $v_{y}u_{x}$ is $uv$ instead of $vu$,
that is, the non--commutativity of dependent variables lies 
in the first part of the symbolic representation indicating the degree of the
expression.

\section{Lax formulation of $(2+1)$--dimensional integrable equations}\label{Sec3}
In this section, we give a short description of construction of
$(2+1)$-dimensional integrable equations from a given scalar Lax operator
based on the well-known Sato approach. For details, see the recent books
\cite{BS01,kp00} and related references in them. 

Consider an $m$-th order formal pseudo-differential operator 
$$
A=a_m D_x^m+a_{m-1} D_x^{m-1}+\cdots +a_0 +a_{-1} D_x^{-1}+\cdots,
\quad m\geq 0,
$$
where the coefficients $a_k$ are functions of $x$
and $D_x^{-1}$ is the formal inverse of the total $x$-derivative $D_x$.
Here the functions $a_k$ take their values in
commutative (for instance complex numbers \(\mathbb{C}\)) or 
noncommutative (for instance the algebra of $n\times n$ matrices) 
associative algebras.

To define the product of two pseudo-differential operators requires 
the action of differential operator $D_x^n$ on a multiplication operator
$f$ (given by a function of $x$): 
\begin{eqnarray*}
D_x^n \circ f=\sum_{i\geq 0} \frac{n(n-1)\cdots (n-i+1)}{i!}(D_x^if) D_x^{n-i}
\end{eqnarray*}
Let the commutator be the bracket on the set of pseudo-differential operators.
Thus, the set of pseudo-differential operators forms a Lie algebra.
For an integer $k<m$, we introduce notations:
\begin{eqnarray*}
&&A_{\geq k}=a_m D_x^m+a_{m-1} D_x^{m-1}+\cdots +a_k D_x^{k}\\
&&A_{<k}=A-A_{\geq k}=a_{k-1} D_x^{k-1}+\cdots
\end{eqnarray*}
When $k\in \{0,1\} $, the Lie algebra decomposes as a direct sum of two
subalgebras in both commutative and noncommutative cases. 
We denote the projections onto these subalgebras by ${\cal P}_{\pm}$.
Such decompositions are naturally related with integrability and
lead to admissible scalar
Lax operators for $(1+1)$-dimensional Lax dynamics (cf. \cite{Bla98} pp. 270
for commutative case), 
namely
\begin{enumerate}
\item[] $k=0: \quad \tilde L=D_x^n+u^{(n-2)}D_x^{n-2}+u^{(n-3)}D_x^{n-3}+\cdots +u^{(0)}, \quad n\geq 2
$;
\item[] $k=1: \quad \tilde L=D_x^n+u^{(n-1)}D_x^{n-1}+u^{(n-2)}D_x^{n-2}+\cdots +u^{(0)}
+D_x^{-1} u^{(-1)}, \quad n\geq 2$;
\end{enumerate}
where $u^{(i)}$ are functions of a spatial variable $x$. 

We know every pseudo-differential operator of order $n>0$ has an $n$-th root.
This allows us to define the fractional powers 
${\tilde L}^{\frac{i}{n}}, i\in \bbbn$.
Let $\tilde B_i={\cal P}_{+}({\tilde L}^{\frac{i}{n}})$.
For each choice of $i$, we introduce a different time variable $t_i$. Thus,
the flows defined by
$$
\frac{\partial {\tilde L}}{\partial {t_i}}=[B_i, \tilde L], 
\quad i\in \bbbn
$$
commute, \cite{gd76}.

This setting up can be generalised into the $(2+1)$-dimensional case as follows:
An $m$-th order pseudo-differential operator of two spatial variables is of
the form
$$
H=-D_y+a_m D_x^m+a_{m-1} D_x^{m-1}+\cdots +a_0 +a_{-1} D_x^{-1}+\cdots,
\quad m\geq 0,
$$
where coefficients $a_k$ are functions of $x, \ y$ and
for an integer $k<m$, we split into
\begin{eqnarray*}
&&H_{\geq k}=a_m D_x^m+a_{m-1} D_x^{m-1}+\cdots +a_k D_x^{k}\\
&&H_{<k}=H-H_{\geq k}=-D_y+a_{k-1} D_x^{k-1}+\cdots
\end{eqnarray*}
Similar to $(1+1)$-dimensional case,
this operator algebra decomposes as a direct sum of two
subalgebras in both commutative and noncommutative cases when $k\in \{0,1\} $. 
The admissible types of scalar Lax operators for the case of $(2+1)$
dimensions are $\tilde L-D_y$, i.e.
\begin{enumerate}
\item[{\bf a}.] $ \quad k=0: \quad L=D_x^n+u^{(n-2)}D_x^{n-2}
+u^{(n-3)}D_x^{n-3}+\cdots +u^{(0)} -D_y, \quad n\geq 2 $;\label{A}
\item[{\bf b}.] $ \quad k=1: \quad L=D_x^n+u^{(n-1)}D_x^{n-1}
+u^{(n-2)}D_x^{n-2}+\cdots +u^{(0)} +D_x^{-1} u^{(-1)}-D_y, \quad n\geq 2$;
\label{B}
\item[{\bf c}.] $ \quad k=1: \quad L=u^{(0)} +D_x^{-1} u^{(-1)}-D_y$;
\label{C}
\end{enumerate}
where $u^{(i)}$ are functions of two spatial variables $x,\ y$. 
We often use $u,v, w, \cdots$ in the examples. 
When $n=1$ for case {\bf b}, we have $L=D_x+u^{(0)} +D_x^{-1} u^{(-1)}-D_y$,
which can be transformed into case {\bf c} by transformation
(\ref{TR}). Therefore, we exclude it from our consideration.

Let $S= D_x+a_0+a_{-1}D_x^{-1}+\cdots$. For any operator $L$ as listed 
in cases {\bf a}, {\bf b} and {\bf c}, the relation
\begin{eqnarray}
&&[S,\ L]:= S L-L S =0,\label{S}
\end{eqnarray}
uniquely determines the operator $S$ by taking the integration 
constants as zero. Furthermore, we have 
$[S^n,\ L]=0$ for any $n\in \bbbn$.
For each choice of $i$, we introduce a different time variable $t_i$ and
define the Lax equation by
\begin{eqnarray}\label{Laxeq}
\frac{\partial L}{\partial {t_i}}=[{S^i}_{\geq k}, L],
\end{eqnarray}
where $k$ is determined by the operator $L$ as listed in cases {\bf a},
{\bf b} and {\bf c}.

\begin{Thm}\label{Th1}
For the operator $S$ uniquely determined as above by (\ref{S}), 
the flows defined by Lax equations (\ref{Laxeq})
commute, that is, $\partial_{ t_j} \partial_{t_i} L=
\partial_{ t_i} \partial_{t_j} L$.
\end{Thm}
{\bf Proof.} Using Lax equations, we have
\begin{eqnarray*}
\partial_{ t_j} \partial_{t_i} L
=[\partial_{ t_j}{S^i}_{\geq k}, L]+
[{S^i}_{\geq k}, \partial_{ t_j}L]
=[\partial_{ t_j}{S^i}_{\geq k}, L]+
[{S^i}_{\geq k}, [{S^j}_{\geq k}, L]]
\end{eqnarray*}
Hence
\begin{eqnarray*}
\partial_{ t_j} \partial_{t_i} L-
\partial_{ t_i} \partial_{t_j} L
=[\partial_{ t_j}{S^i}_{\geq k}
-\partial_{ t_i}{S^j}_{\geq k}-[{S^j}_{\geq k}, {S^i}_{\geq k}],L].
\end{eqnarray*}
Meanwhile, from formula (\ref{S}), we get
\begin{eqnarray*}
[\partial_{t_i} S, \ L]=-[S, \partial_{t_i}L] 
=-[S, [{S^i}_{\geq k}, L]]
=[[{S^i}_{\geq k},S], L]
\end{eqnarray*}
implying 
\begin{eqnarray}\label{Sato}
\partial_{t_i} S=[{S^i}_{\geq k},S].
\end{eqnarray}
Thus 
\begin{eqnarray*}
&&\partial_{ t_j}{S^i}_{\geq k}
-\partial_{ t_i}{S^j}_{\geq k}-[{S^j}_{\geq k}, {S^i}_{\geq k}]
\\&&
=(\partial_{ t_j}{S^i})_{\geq k}
-(\partial_{ t_i}{S^j})_{\geq k}-[{S^j}_{\geq k}, {S^i}_{\geq k}]
=[{S^j}_{\geq k}, S^i]_{\geq k} -[{S^i}_{\geq k}, S^j]_{\geq k}
-[{S^j}_{\geq k}, {S^i}_{\geq k}]
\\&&
=-[{S^j}_{<k}, {S^i}_{\geq k}+{S^i}_{<k}]_{\geq k} 
-[{S^i}_{\geq k}, {S^j}_{\geq k}+{S^j}_{<k}]_{\geq k}
-[{S^j}_{\geq k}, {S^i}_{\geq k}]
=-[{S^j}_{<k}, {S^i}_{<k}]_{\geq k}
\\&&
=0.
\end{eqnarray*}
This leads to
$\partial_{ t_j} \partial_{t_i} L=
\partial_{ t_i} \partial_{t_j} L$.
$\diamond$

\begin{Rem}\label{Rem1}
In the commutative case, there is one more admissible type of scalar 
Lax operator, that is,
$$ k=2: \quad L={u^{(n)}}^n D_x^n+u^{(n-1)}D_x^{n-1}+u^{(n-2)}
D_x^{n-2}+\cdots +u^{(0)} +D_x^{-1} u^{(-1)}+D_x^{-2} u^{(-2)}-D_y.$$
Theorem \ref{Th1} is also valid for this type by taking
$S=u^{(n)} D_x+a_0+a_{-1}D_x^{-1}+\cdots$ instead.
\end{Rem}

\begin{Ex}\label{Ex4}
Consider the Lax operator $L=u^2 D_x^2 +v D_x +w + D_x^{-1} p +D_x^{-2} q -D_y$,
where all dependent variables take their values in $\mathbb{C}$.
Using formula (\ref{S}), we have $S=u D_x+a_0+a_{-1}D_x^{-1}+\cdots$ with
\begin{eqnarray*}
&&a_0=-\frac{1}{2}(u_x-\frac{v}{u}+\Theta\frac{1}{u})\\
&&a_{-1}=
\frac{1}{8}(2 u_{xx}-\frac{u_x^2}{u}-\frac{2v_x}{u}+\frac{4 u_x v}{u^2}
-\frac{v^2}{u^3}-\frac{4u_y}{u^2}+\frac{2}{u}D_x^{-1}D_y\frac{v}{u^2}
+\frac{4w}{u}-\frac{2}{u}\Theta(\frac{1}{u}\Theta\frac{1}{u})
+\frac{1}{u}(\Theta\frac{1}{u})^2).
\end{eqnarray*}
and ${S^1}_{\geq 2}=0$,\quad  ${S^2}_{\geq 2}=u^2 D_x^2$,\quad
${S^3}_{\geq 2}=u^3D_x^3 +3 u^2(u_x+a_0)D_x^2$. By the Lax equation, we obtain
\begin{eqnarray*}
&&\left\{\begin{array}{l}
u_{t_3}=\frac{1}{4}(u^3 u_{xxx}-6 u_y \Theta\frac{1}{u}
-3 u (\Theta\frac{1}{u})_y+3 u^2 v_{xx}+3 u^2 w_x
+3 v_y+3v v_x -3 \frac{u_x}{u} v^2-6 (uv)_x \Theta\frac{1}{u})
\\
v_{t_3}=u^3v_{xxx}+3u^2(w_{xx}+u p_x+u_{x}p+(v_{xx}+2 w_x)(u_x+a_0))
\\
w_{t_3}=u^3 w_{xxx}+3u^2(w_{xx}+p_x)(u_x+a_0)+3p(u^2a_0)_x
+3u^2 (uq_x-2u_x p_x+2u_xq)
\\
p_{t_3}=(u^3 p)_{xxx}-3\left([u^2 (u_x p+a_0 p+q)]_{xx}+2 [u^2 q (u_x+a_0)]_x\right)
\\
q_{t_3}=(u^3 q)_{xxx}-3 [u^2 q(u_x+a_0)]_{xx}
\end{array}\right.
\end{eqnarray*}
Notice that the reduction of $v=w=p=q=0$ leads to well-known 
$(2+1)$-Harry-Dym equation
$$
u_{t_3}=\frac{1}{4}(u^3 u_{xxx}-6 u_y \Theta\frac{1}{u}
-3 u (\Theta\frac{1}{u})_y ) \ .
$$
\end{Ex}

\section{Lax formulation in symbolic representation}\label{Sec4}
In this section, we carry out the formalism of section \ref{Sec3} in symbolic
form. 
We first consider noncommutative dependent variables to simplify
our formulae.
The strategy for the commutative case is to do the calculation
as much as possible by treating it as noncommutative and only do
symmetrisation at the last stage to get the uniqueness of the symbolic
representation since the symmetrisation complicates the calculation dramatically.

Let us assign the symbols $\xi^{(i)}, \eta^{(i)}$ for dependent variable $u^{(i)}$ and the symbol $Y$ for the operator $D_y$.
The symbolic representations of the admissible scalar Lax operators are
\begin{enumerate}
\item[{\bf a}.] $ \quad k=0: \quad \hat L=X^n-Y+u^{(n-2)}X^{n-2}
+u^{(n-3)}X^{n-3} +\cdots +u^{(0)},\quad n\geq 2 $;
\item[{\bf b}.] $ \quad k=1: \quad \hat L=X^n-Y+u^{(n-1)}X^{n-1}
+u^{(n-2)}X^{n-2}+\cdots +u^{(0)}
+ u^{(-1)}\frac{1}{X+\xi^{(-1)}},\quad n\geq 2 $;
\item[{\bf c}.] $ \quad k=1: \quad 
\hat L=-Y+u^{(0)}+u^{(-1)}\frac{1}{X+\xi^{(-1)}}$;
\end{enumerate}
We first treat case {\bf a}. 
It is convenient to consider formal series in the form
\begin{eqnarray}
S&=&X+\sum_{i=0}^{n-2} u^{(i)} a_1^{(i)}(\xi_1^{(i)},\eta_1^{(i)},X)
+\sum_{i_1=0}^{n-2} \sum_{i_2=0}^{n-2}
u^{(i_1)} u^{(i_2)}a_2^{(i_1 i_2)}(\xi_{j_1}^{(i_1)},\eta_{j_1}^{(i_1)},
\xi_{j_2}^{(i_2)},\eta_{j_2}^{(i_2)},X)
+\cdots, n\geq 2
\label{SA}
\end{eqnarray}
where $a_i$ are functions of their specific arguments
and $j_k$ is defined by the number of $i_k$ in the list of $[i_1, i_2, \cdots,
i_l]$, which implies that $j_1=1$ and $j_k\geq 1$, $k=1,2,\cdots ,l$. 
For example, when $i_1=i_2$, the arguments of function $a_2^{i_1i_1}$  
are $\xi_{1}^{(i_1)}$, $\eta_{1}^{(i_1)}$,
$\xi_{2}^{(i_1)}$, $\eta_{2}^{(i_1)}$ and $X$.

It is easy to check that
\begin{eqnarray}
[X^n-Y,\ \phi(\xi_{j_1}^{(i_1)},\eta_{j_1}^{(i_1)},
\cdots,\xi_{j_l}^{(i_l)},\eta_{j_l}^{(i_l)},X)]
=N_l(\xi_{j_1}^{(i_1)},\eta_{j_1}^{(i_1)},
\cdots,\xi_{j_l}^{(i_l)},\eta_{j_l}^{(i_l)},X) \phi,
\label{linear}
\end{eqnarray}
where the polynomial $N_l$ is defined by
\begin{eqnarray}
 N_l(x_1,y_1,x_2,y_2,\ldots, x_l,y_l;X)=(\sum_{i=1}^l x_i +X)^n-X^n-\sum_{i=1}^l y_i\label{dem}
\end{eqnarray}

\begin{Pro}\label{Pro1}
For any operator $L$ in case {\bf a}, if the formal series (\ref{SA}) 
satisfies the relation 
$[S,\ L]=0$ (cf. (\ref{S})),
we have for $l\geq 1$,
\begin{eqnarray}
a_{l}^{(i_1i_2\cdots i_{l})}=
a_l(\xi_{j_1}^{(i_1)},\eta_{j_1}^{(i_1)},\cdots,\xi_{j_l}^{(i_l)},\eta_{j_l}^{(i_l)},X)=
\prod_{r=1}^{l}(X+\sum_{s=r+1}^{l}\xi_{j_s}^{(i_s)})^{i_r}
b_{l}(\xi_{j_1}^{(i_1)},\eta_{j_1}^{(i_1)},\cdots,\xi_{j_l}^{(i_l)},
\eta_{j_l}^{(i_l)},X)
\label{Al},
\end{eqnarray}
where the superindex $i_s\in \{0,1,2,\cdots, n-2\}$
and the subindex $j_k$ is defined by the number of $i_k$ in
the list of $[i_1, i_2, \cdots, i_k]$. 
The function $b_l$, $l\geq1$, is defined by
\begin{eqnarray}
&&b_l(\xi_{j_1}^{(i_1)},\eta_{j_1}^{(i_1)},\cdots,\xi_{j_l}^{(i_l)},\eta_{j_l}^{(i_l)},X)
=\frac{c_l(\xi_{j_1}^{(i_1)},\eta_{j_1}^{(i_1)},\cdots,\xi_{j_l}^{(i_l)},\eta_{j_l}^{(i_l)},X)}
{N_l(\xi_{j_1}^{(i_1)},\eta_{j_1}^{(i_1)},\cdots,\xi_{j_l}^{(i_l)},\eta_{j_l}^{(i_l)},X)}.
\label{Bf}
\end{eqnarray}
with 
\begin{eqnarray}
&&c_l(\xi_{j_1}^{(i_1)},\eta_{j_1}^{(i_1)},\cdots,\xi_{j_l}^{(i_l)},\eta_{j_l}^{(i_l)},X)=
\nonumber\\&=&
b_{l-1}(\xi_{j_1}^{(i_1)},\eta_{j_1}^{(i_1)}\cdots,\xi_{j_{l-1}}^{(i_{l-1})},
\eta_{j_{l-1}}^{(i_{l-1})}, X+\xi_{j_{l}}^{(i_l)})
-b_{l-1}(\xi_{j_2}^{(i_2)},\eta_{j_2}^{(i_2)},\cdots, \xi_{j_{l}}^{(i_l)},\eta_{j_{l}}^{(i_l)},X), \quad l>1
\label{Cf}
\end{eqnarray}
and the initial function
$c_1(\xi_1^{(i)},\eta_1^{(i)},X)=\xi_1^{(i)}$.
\end{Pro}
{\bf Proof.} We compute $[L,\ S]$ symbolically and collect the coefficients
of each algebraically independent monomial of the dependent variables.
For the linear terms, we have for any $i\in \{0, \cdots, n-2\}$,
\begin{eqnarray}
a_1^{(i)}(\xi_1^{(i)},\eta_1^{(i)},X)
=\frac{\xi_1^{(i)} X^i}{N_1(\xi_1^{(i)},\eta_1^{(i)},X)}
= b_1(\xi_1^{(i)},\eta_1^{(i)},X) X^i.
\label{A1}
\end{eqnarray}
For the quadratic terms,  when $i\neq j$, we need to compute
\begin{eqnarray*}
A^{ij}:&=&-X^i\circ a_1^{(j)}(\xi_1^{(j)},\eta_1^{(j)},X)
+a_1^{(i)}(\xi_1^{(i)},\eta_1^{(i)},X)\circ X^j
\\&=&
-\frac{(X+\xi_1^{(j)})^i\xi_1^{(j)} X^j}{N_1(\xi_1^{(j)},\eta_1^{(j)},X)}
+\frac{\xi_1^{(i)} (X+\xi_1^{(j)})^i X^j}{N_1(\xi_1^{(i)},\eta_1^{(i)},X+\xi_1^{(j)})}
\\&=&
(X+\xi_1^{(j)})^i X^j (\frac{\xi_1^{(i)}}{N_1(\xi_1^{(i)},\eta_1^{(i)},X+\xi_1^{(j)})}
-\frac{\xi_1^{(j)}}{N_1(\xi_1^{(j)},\eta_1^{(j)},X)})
\\&=&
(X+\xi_1^{(j)})^i X^j (b_1(\xi_1^{(i)},\eta_1^{(i)},X+\xi_1^{(j)})
-b_1(\xi_1^{(j)},\eta_1^{(j)},X))
\\&=&
(X+\xi_1^{(j)})^i X^j c_2(\xi_1^{(i)},\eta_1^{(i)},\xi_1^{(j)},\eta_1^{(j)},X)
\end{eqnarray*}
and when $i=j$, we have 
\begin{eqnarray*}
A^{ii}:&=&-[X^i,\ a_1^{(i)}(\xi_1^{(i)},\eta_1^{(i)},X)]
=-X^i\circ a_1^{(i)}(\xi_1^{(i)},\eta_1^{(i)},X)
+a_1^{(i)}(\xi_1^{(i)},\eta_1^{(i)},X)\circ X^i
\\&=&
-\frac{(X+\xi_2^{(i)})^i\xi_2^{(i)} X^i}{N_1(\xi_2^{(i)},\eta_2^{(i)},X)}
+\frac{\xi_1^{(i)} (X+\xi_2^{(i)})^i X^i}{N_1(\xi_1^{(i)},\eta_1^{(i)},X+\xi_2^{(i)})}.
\end{eqnarray*}
This can be obtained by substituting $\xi_1^{(j)}=\xi_2^{(i)}$ and
$\eta_1^{(j)}=\eta_2^{(i)}$ in $A^{ij}$.
It follows that 
\begin{eqnarray}
&&a_2^{(i_1 i_2)}(\xi_{j_1}^{(i_1)},\eta_{j_1}^{(i_1)},\xi_{j_2}^{(i_2)},\eta_{j_2}^{(i_2)},X)
=\frac{A^{i_1 i_2}}{N_2(\xi_{j_1}^{(i_1)},\eta_{j_1}^{(i_1)},\xi_{j_2}^{(i_2)},\eta_{j_2}^{(i_2)},X)}\nonumber\\
&=&(X+\xi_{j_2}^{(i_2)})^{i_1} X^{i_2} 
b_2(\xi_{j_1}^{(i_1)},\eta_{j_1}^{(i_1)},\xi_{j_2}^{(i_2)},\eta_{j_2}^{(i_2)},X).
\label{A2n}
\end{eqnarray}
Assume that formula (\ref{Al}) is valid for degree $l-1$.
Let us compute $a_{l}^{(i_1i_2\cdots i_l)}$ in the same way as
computing the quadratic terms. Then we have
\begin{eqnarray*}
&&N_l(\xi_{j_1}^{(i_1)},\eta_{j_1}^{(i_1)},\cdots,\xi_{j_l}^{(i_l)},\eta_{j_l}^{(i_l)},X)
a_{l}^{(i_1i_2\cdots i_l)}
=-X^{i_1}\circ a_{l-1}^{(i_2\cdots i_l)}
+a_{l-1}^{(i_1i_2\cdots i_{l-1})}\circ X^{i_l}
\\&=&
-(X+\sum_{p=2}^{l}\xi_{j_p}^{(i_p)})^{i_1}
\prod_{r=2}^{l}(X+\sum_{s=r+1}^{l}\xi_{j_s}^{(i_s)})^{i_r}
b_{l-1}(\xi_{j_2}^{(i_2)},\eta_{j_2}^{(i_2)},\cdots,\xi_{j_l}^{(i_{l})},
\eta_{j_l}^{(i_{l})},X)
\\&&
+\prod_{r=1}^{l-1}(X+\sum_{s=r+1}^{l}\xi_{j_s}^{(i_s)})^{i_r} X^{i_l}
b_{l-1}(\xi_{j_1}^{(i_1)},\eta_{j_1}^{(i_1)},\cdots,\xi_{j_{l-1}}^{(i_{l-1})},
\eta_{j_{l-1}}^{(i_{l-1})},X+\xi_{j_{l}}^{(i_{l})})
\\&=&
\prod_{r=1}^{l}(X+\sum_{s=r+1}^{l}\xi_{j_s}^{(i_s)})^{i_r}
(b_{l-1}(\xi_{j_1}^{(i_1)},\eta_{j_1}^{(i_1)},\cdots,\xi_{j_{l-1}}^{(i_{l-1})},
\eta_{j_{l-1}}^{(i_{l-1})},X+\xi_{j_{l}}^{(i_{l})})
-b_{l-1}(\xi_{j_2}^{(i_2)},\eta_{j_2}^{(i_2)},\cdots,\xi_{j_l}^{(i_{l})},
\eta_{j_l}^{(i_{l})},X)).
\\&=&
\prod_{r=1}^{l}(X+\sum_{s=r+1}^{l}\xi_{j_s}^{(i_s)})^{i_r}
c_{l}(\xi_{j_1}^{(i_1)},\eta_{j_1}^{(i_1)},\cdots,
\xi_{j_l}^{(i_{l})}, \eta_{j_l}^{(i_{l})},X).
\end{eqnarray*}
According to definition (\ref{Bf}), we proved that formula (\ref{Al}) is valid
for degree $l$.
$\diamond$

Similarly, we prove the following result for any operator $L$ in cases {\bf b}
and {\bf c}.
\begin{Pro}\label{Pro2}
For any operator $L$ in case {\bf b}, consider 
a formal series of the form 
\begin{eqnarray}\label{Sk1}
S=
X+\sum_{i=-1}^{n-1} u^{(i)} a_1^{(i)}(\xi_1^{(i)},\eta_1^{(i)},X)
+\sum_{i_1,i_2=-1}^{n-1}u^{(i_1)} u^{(i_2)}a_2^{(i_1 i_2)}(\xi_{j_1}^{(i_1)},\eta_{j_1}^{(i_1)},
\xi_{j_2}^{(i_2)},\eta_{j_2}^{(i_2)},X)
+\cdots,  n\geq 2
\end{eqnarray}
Here $a_i$ are functions of their specific arguments,
the superindex $i_s\in \{-1,0,1,\cdots, n-1\}$
and the subindex $j_k$ is defined by the number of $i_k$ 
in the list of $[i_1, i_2, \cdots, i_k]$.
If it satisfies the relation 
$[S,\ L]=0$ (cf. (\ref{S})),
we have for $l\geq 1$,
\begin{eqnarray}
a_{l}^{(i_1i_2\cdots i_{l})}=
a_l(\xi_{j_1}^{(i_1)},\eta_{j_1}^{(i_1)},\cdots,\xi_{j_l}^{(i_l)},\eta_{j_l}^{(i_l)},X)=
\prod_{r=1}^{l}(h(i_r)+\sum_{s=r+1}^{l}\xi_{j_s}^{(i_s)})^{i_r}
b_{l}(\xi_{j_1}^{(i_1)},\eta_{j_1}^{(i_1)},\cdots,\xi_{j_l}^{(i_{l})},
\eta_{j_l}^{(i_{l})},X),
\label{ABl}
\end{eqnarray}
where $h(i_r)=X+\xi_{j_r}^{(-1)}$ if $i_r=-1$, otherwise $h(i_r)=X$.
The functions $b_l$ are defined in 
Proposition \ref{Pro1}, cf. formula (\ref{Bf}) and (\ref{Cf}).
\end{Pro}
Notice that for operator $L$ listed in case {\bf c},  we have 
$N_l(x_1,y_1,x_2,y_2,\ldots, x_l,y_l;X)=-\sum_{i=1}^l y_i$, cf. (\ref{linear}),
which corresponds to the action of total $y$-derivative  $D_y$. Thus
\begin{Pro}\label{Pro0}
For operator $L$ listed in case {\bf c},
if a formal series of the form 
\begin{eqnarray}\label{Sk0}
S=
X+\sum_{i=-1}^{0} u^{(i)} a_1^{(i)}(\xi_1^{(i)},\eta_1^{(i)},X)
+\sum_{i_1,i_2=-1}^{0} u^{(i_1)} u^{(i_2)}a_2^{(i_1 i_2)}(\xi_{j_1}^{(i_1)},\eta_{j_1}^{(i_1)},
\xi_{j_2}^{(i_2)},\eta_{j_2}^{(i_2)},X)
+\cdots,
\end{eqnarray}
where the superindex $i_s\in \{-1,0\}$
and the subindex $j_k$ is defined by the number of $i_k$ 
in the list of $[i_1, i_2, \cdots, i_k]$,
satisfies the relation 
$[S,\ L]=0$ (cf. (\ref{S})),
we have for $l\geq 1$, the functions $a_l$
are defined in Proposition \ref{Pro2}, cf. formula (\ref{ABl})
with $N_l(x_1,y_1,x_2,y_2,\ldots, x_l,y_l;X)=-\sum_{i=1}^l y_i$.
\end{Pro}

To construct the hierarchy of the Lax equations
we need to expand the coefficients of operator (\ref{SA}) or (\ref{Sk1})
or (\ref{Sk0}) at $X=\infty$ and truncate at the required degree.
\begin{Def}
We say that a function $h(x_1,y_1,x_2,y_2,\cdots, x_l,y_l,X)$ 
is $k$th (quasi-local) polynomial
if the first $k$ coefficients of its expansion at $X= \infty$
are symbols of (quasi-local) polynomials.
If a function is $k$th (quasi-local) polynomial for any $k$, 
we say it is (quasi-local) polynomial.
\end{Def}

When $n \geq 2$, that is, operator $L$ in cases {\bf a} and {\bf b}, 
the expansion of $N_l(x_1,y_1,x_2,y_2,\ldots, x_l, y_l;X)$
at $X= \infty$ is of the form
\begin{eqnarray*}
&&N_l(x_1,y_1,x_2,y_2,\ldots, x_l, y_l;X)^{-1}
\\&=&
\frac{1}{n X^{n-1} (\sum_{i=1}^l x_i)}
\sum_{j\geq 0}[\frac{\sum_{i=1}^l y_i}{n X^{n-1} (\sum_{i=1}^l x_i)}
-\frac{1}{n}\sum_{k=0}^{n-2} \left( \begin{array}{l} n\\k \end{array}\right) \
X^{k+1-n} (\sum_{i=0}^l x_i)^{n-k-1}]^j,\quad n\geq 2.
\end{eqnarray*}
For operator $L$ listed in case {\bf c}, we know
$N_l(x_1,y_1,x_2,y_2,\ldots, x_l,y_l;X)=-\sum_{i=1}^l y_i$.
Therefore, if we want to prove that the coefficients of operators (\ref{SA}),
(\ref{Sk1}) and (\ref{Sk0}), i.e. the functions $a_l$,  are quasi-local,
we need to show that the functions $c_l$
can be split into the sum of the image of $D_x$ and
the image of $D_y$.
It is clear from formula (\ref{A1}) that 
$a_1^{(i)}(\xi_1^{(i)},\eta_1^{(i)},X)$ are quasi-local since 
we have $c_1(\xi_1^{(i)},\eta_1^{(i)},X)=\xi_1^{(i)}$.
Now we concentrate on the cases when $l>1$.
\begin{Pro}\label{Pro3}
The functions
$c_l(\xi_{j_1}^{(i_1)},\eta_{j_1}^{(i_1)},\cdots,\xi_{j_l}^{(i_l)},\eta_{j_l}^{(i_l)},X)$ for $l>1$ vanish after substitution
$$\xi_{j_1}^{(i_1)}=-\xi_{j_2}^{(i_2)}-\cdots -\xi_{j_{l-1}}^{(i_{l-1})}\quad \mbox {and\quad}
\eta_{j_1}^{(i_1)}=-\eta_{j_2}^{(i_2)}-\cdots -\eta_{j_{l-1}}^{(i_{l-1})}. $$
\end{Pro}

We give its proof in the Appendix. In fact, this 
proposition does not lead to our intended conclusion that
$b_l$ and thus $a_l$ are quasi-local since the objects 
are rational, not polynomial. For example, the expression 
$u^2(\frac{\eta_2}{\xi_2}-\frac{\eta_1}{\xi_1})$ representing
$ u \Theta u-(\Theta u) u$ satisfies the above proposition. However,
we can't write $u \Theta u-(\Theta u) u=D_x f_1+D_y f_2$,
where both $f_1$ and $f_2 $ are in ${\cal U} (\Theta)$.

We first have a close look at the quadratic terms.
For any operator $L$ in cases {\bf a} and {\bf b},
let $\eta_{j_r}^{i_r}=\theta_{j_r}^{i_r} \xi_{j_r}^{i_r}$. Then
the function $c_2(\xi_{j_1}^{i_1},\eta_{j_1}^{i_1},
\xi_{j_2}^{i_2},\eta_{j_2}^{i_2},X)$, where $\xi_{j_1}^{i_1}\neq 0$
and $\xi_{j_2}^{i_2}\neq 0$,
is a polynomial of $\theta_{j_r}^{i_r},\xi_{j_r}^{i_r}$,
$r=1,2$. 
Notice  that 
\begin{eqnarray}
&&
(\eta_{j_1}^{i_1}+\eta_{j_2}^{i_2})
=\theta_{j_1}^{i_1}\xi_{j_1}^{i_1}
+\theta_{j_2}^{i_2} \xi_{j_2}^{i_2}
=(\theta_{j_1}^{i_1}-\theta_{j_2}^{i_2})\xi_{j_1}^{i_1}
+\theta_{j_2}^{i_2} (\xi_{j_1}^{i_1}+\xi_{j_2}^{i_2})
=\theta_{j_1}^{i_1} (\xi_{j_1}^{i_1}+\xi_{j_2}^{i_2})
-(\theta_{j_1}^{i_1}-\theta_{j_2}^{i_2})\xi_{j_2}^{i_2}.
\label{Var}
\end{eqnarray}
The affine variety (cf. \cite{mr97h:13024}) defined by 
$\xi_{j_1}^{i_1}+\xi_{j_2}^{i_2}$
and $\eta_{j_1}^{i_1}+\eta_{j_2}^{i_2}$ is
\begin{eqnarray*}
&&V(\xi_{j_1}^{i_1}+\xi_{j_2}^{i_2}, \eta_{j_1}^{i_1}+\eta_{j_2}^{i_2})
=V(\xi_{j_1}^{i_1}+\xi_{j_2}^{i_2}, 
(\theta_{j_1}^{i_1}-\theta_{j_2}^{i_2})\xi_{j_1}^{i_1},
(\theta_{j_1}^{i_1}-\theta_{j_2}^{i_2})\xi_{j_2}^{i_2})
\\&=&
V(\xi_{j_1}^{i_1}+\xi_{j_2}^{i_2}, \theta_{j_1}^{i_1}-\theta_{j_2}^{i_2})
\cup V(\xi_{j_1}^{i_1},\xi_{j_2}^{i_2})
\end{eqnarray*}

From Proposition \ref{Pro3}, it can be written as
\begin{eqnarray*}
&&c_2(\xi_{j_1}^{i_1},\theta_{j_1}^{i_1}\xi_{j_1}^{i_1},
\xi_{j_2}^{i_2},\theta_{j_2}^{i_2} \xi_{j_2}^{i_2},X)
=(\xi_{j_1}^{i_1}+\xi_{j_2}^{i_2})f_1
+(\theta_{j_1}^{i_1}-\theta_{j_2}^{i_2})f_2,
\end{eqnarray*}
where both $f_1$ and $f_2$ are polynomials of 
$\theta_{j_r}^{i_r},\xi_{j_r}^{i_r}$, $r=1,2$.
From formula (\ref{Var}), the part that is not in the image
of $D_x$ or $D_y$ only depends
on $\theta_{j_r}^{i_r}$, $r=1,2$. 

For operator $L$  listed in case {\bf c}, since the function
$N_l(x_1,y_1,\cdots, x_l,y_l,X)=-\sum_{i=1}^l y_i$, cf. Proposition \ref{Pro0},
we let $\xi_{j_r}^{i_r}=\theta_{j_r}^{i_r} \eta_{j_r}^{i_r}$ instead. 
Then the function $c_2(\xi_{j_1}^{i_1},\eta_{j_1}^{i_1},
\xi_{j_2}^{i_2},\eta_{j_2}^{i_2},X)$, where $\eta_{j_1}^{i_1}\neq 0$
and $\eta_{j_2}^{i_2}\neq 0$,
is a polynomial of $\theta_{j_r}^{i_r},\xi_{j_r}^{i_r}$,
$r=1,2$. The above discussion is valid.

Setting $\xi_{j_1}^{i_1}=\xi_{j_2}^{i_2}=0$,
we obtain  the function $c_2(\xi_{j_1}^{i_1},\eta_{j_1}^{i_1},
\xi_{j_2}^{i_2},\eta_{j_2}^{i_2},X)$ equals
\begin{eqnarray}
\frac{1}{n X^{n-1}-\theta_{j_1}^{i_1}}
-\frac{1}{n X^{n-1}-\theta_{j_2}^{i_2}}
\quad \mbox{for cases {\bf a} and {\bf b}};
\quad
\theta_{j_2}^{i_2} -\theta_{j_1}^{i_1}
\quad \mbox{for case {\bf c}},
 \label{Cqu}
\end{eqnarray}
which is zero after we symmetrise it with the permutation group $\Sigma_2$.
This implies that the functions $b_2$ and thus $a_2$ are quasi-local 
in the commutative case.

\subsection{The commutative case}
We first give the formulae to compute the high degree terms of
operator $S$ for the commutative case directly from the formulae we obtained 
in Proposition \ref{Pro1}, \ref{Pro2} and \ref{Pro0}.

We denote the list $(i_1, i_2, \cdots, i_l)$ by $I$ and 
$I_1=(i_{s_1}, i_{s_2},\cdots, i_{s_r})$, 
where $1\leq s_1<s_2<\cdots <s_r\leq l$ and $1\leq r\leq l-1$.
We use $I_2=(i_{p_1}, i_{p_2},\cdots, i_{p_{l-r}})$ 
to denote the list by removing the elements of list $I_1$
from list $I$. For any element $i_r$ 
in the set $\{i_1, i_2, \cdots, i_l\}$, we denote the number of $i_r$ 
in the list $I$ by $\#i_r$. 
Then the coefficient of 
$\prod_{i_r\in \{i_1 i_2 \cdots i_l\}} (u^{(i_r)})^{\#i_r}$  in 
operator $s$, cf. formula (\ref{SA}), (\ref{Sk1}) and (\ref{Sk0}),
can be computed by
\begin{eqnarray}
\frac{1}{\prod_{i_r\in \{i_1 i_2 \cdots i_l\}} (\#i_r)!} 
\sum_{\sigma\in \Sigma_l}
a_{l}(\xi_{j_{\sigma(1)}}^{(i_{\sigma(1)})},\eta_{j_{\sigma(1)}}^{(i_{\sigma(1)})},
\cdots,\xi_{j_{\sigma(l)}}^{(i_{\sigma(l)})}, \eta_{j_{\sigma(l)}}^{(i_{\sigma(l)})},X).
\label{CN}
\end{eqnarray}
We denote it by
\begin{eqnarray*}
\frac{d_l^{i_1 i_2\cdots i_l}}{N_l(\xi_{j_1}^{(i_1)},\eta_{j_1}^{(i_1)},\cdots,\xi_{j_l}^{(i_l)},\eta_{j_l}^{(i_l)},X)}.
\end{eqnarray*}
In particular, for any operator $L$ in case {\bf a}, we have
\begin{eqnarray}
d_l^{i_1 i_2\cdots i_l}
=\frac{1}{\prod_{i_r\in \{i_1 i_2 \cdots i_l\}} (\#i_r)!} 
\sum_{\sigma\in \Sigma_l}
\prod_{r=1}^{l}(X+\sum_{s=r+1}^{l}\xi_{j_{\sigma(s)}}^{(i_{\sigma(s)})})
^{i_{\sigma(r)}}
c_{l}(\xi_{j_{\sigma(1)}}^{(i_{\sigma(1)})},\eta_{j_{\sigma(1)}}^{(i_{\sigma(1)})},
\cdots,\xi_{j_{\sigma(l)}}^{(i_{\sigma(l)})}, \eta_{j_{\sigma(l)}}^{(i_{\sigma(l)})},X).
\label{CAl}
\end{eqnarray}

For cases {\bf b} and {\bf c}, the expression $(X+p)^{-1}$, 
where $p=\xi_{j_r}^{(-1)}+\sum_{s=r+1}^{l}\xi_{j_s}^{(i_s)}$ 
appears in formula (\ref{ABl}). Its expansion at $X= \infty$
is of the form
\begin{eqnarray}
(X+p)^{-1}=\frac{1}{X}\sum_{j=0}^{+\infty} 
(-\frac{p}{X})^s,
\label{E-1}
\end{eqnarray}
whose coefficients are polynomials in $p$. This does not affect the proof 
and result.  For case {\bf c}, we need to exchange the symbols
$\xi_{j_r}^{i_r}$ and $\eta_{j_r}^{i_r}$ in the proof as the discussion we did 
for quadratic terms.
Therefore, we'll only give the proof for this case.

The immediate consequence of the previous discussion on the quadratic terms
is the following statement.
\begin{Pro}\label{Pro4}
All quadratic terms in operator $S$ are quasi-local.
\end{Pro}

In fact, we can prove this statement is valid for all $l>0$.
\begin{Thm}\label{Th2}
Every term in the operator $S$, 
cf. formula (\ref{SA}), (\ref{Sk1}) and (\ref{Sk0}),
obtained via (\ref{CN}) is quasi-local.
\end{Thm}
{\bf Proof.} 
We proved the linear and quadratic terms in $S$ are quasi-local, that is,
the statement is true for $l=1,2$. 
Assume it is also true for $l-1$. By induction,
the function $d_l^{i_1 i_2\cdots i_l}$, cf. (\ref{CAl}),
is polynomial of $\xi_{j_r}^{(i_r)}$
and $\theta_{{I_1}}=
\frac{\eta_{j_{s_1}}^{(i_{s_1})}+\cdots \eta_{j_{s_r}}^{(i_{s_r})}}
{\xi_{j_{s_1}}^{(i_{s_1})}+\cdots \xi_{j_{s_r}}^{(i_{s_r})}}$,
$r=1,2,\cdots,l-1$,
which obviously satisfies Proposition \ref{Pro3}.
This implies that the function $d_l^{i_1 i_2\cdots i_l}$
is the sum of the parts with factors
$\xi_{j_1}^{(i_1)}+\xi_{j_2}^{(i_2)}+\cdots +\xi_{j_l}^{(i_l)}$
or $\theta_{I_1}-\theta_{I_2}$ using the argument of the affine variety.
From the recurrent relation in 
Proposition \ref{Pro1}, we can compute the factor for any 
$\theta_{I_1}-\theta_{I_2}$ (without symmetrisation), which equals
\begin{eqnarray*}
&&\prod_{q=1}^r (X+\xi_{j_{s_{q+1}}}^{(i_{s_{q+1}})}+\cdots 
+\xi_{j_{s_r}}^{(i_{s_r})}
+\xi_{j_{p_1}}^{(i_{p_1})}+\cdots +\xi_{j_{p_{l-r}}}^{(i_{p_{l-r}})})^{i_{s_q}}
\prod_{q=1}^{l-r} (X+\xi_{j_{p_{q+1}}}^{(i_{p_{q+1}})}+\cdots 
+\xi_{j_{p_{l-r}}}^{(i_{p_{l-r}})})^{i_{p_q}}
\\
&&-\prod_{q=1}^{l-r} (X+\xi_{j_{p_{q+1}}}^{(i_{p_{q+1}})}+\cdots 
+\xi_{j_{p_{l-r}}}^{(i_{p_{l-r}})}
+\xi_{j_{s_1}}^{(i_{s_1})}+\cdots +\xi_{j_{s_{r}}}^{(i_{s_{r}})})^{i_{p_q}}
\prod_{q=1}^{r} (X+\xi_{j_{s_{q+1}}}^{(i_{s_{q+1}})}+\cdots 
+\xi_{j_{s_{r}}}^{(i_{s_{r}})})^{i_{s_q}}.
\end{eqnarray*}
This expression is zero under the substitution 
$\Xi_{I_2}=\xi_{j_{p_1}}^{(i_{p_1})}+\cdots +\xi_{j_{p_{l-r}}}^{(i_{p_{l-r}})}=0$
and $\Xi_{I_1}=\xi_{j_{s_1}}^{(i_{s_1})}+\cdots +\xi_{j_{s_{r}}}^{(i_{s_{r}})}=0$
implying we have either
$\Xi_{I_2} (\theta_{I_1}-\theta_{I_2}) $
or $\Xi_{I_1} (\theta_{I_1}-\theta_{I_2}) $ as a factor.
We know
\begin{eqnarray*}
&&\Xi_{I_2} (\theta_{I_1}-\theta_{I_2}) 
=\theta_{I_1} (\xi_{j_1}^{(i_1)}+\cdots +\xi_{j_l}^{(i_l)})
-(\eta_{j_1}^{(i_1)}+\cdots +\eta_{j_l}^{(i_l)});
\\
&&\Xi_{I_1} (\theta_{I_1}-\theta_{I_2})
=(\eta_{j_1}^{(i_1)}+\cdots +\eta_{j_l}^{(i_l)})
-\theta_{I_2} (\xi_{j_1}^{(i_1)}+\cdots +\xi_{j_l}^{(i_l)}).
\end{eqnarray*}
Thus the $l$-th degree term of $S$ is quasi-local.
$\diamond$

This theorem implies that every terms in $S^n$ is quasi-local. From Theorem 
\ref{Th1}, we have
\begin{Thm}
The hierarchies of ($2+1$)-integrable equation with scalar Lax operators
are quasi-local.
\end{Thm}
As mentioned in Remark \ref{Rem1}, there is one more type of scalar Lax
operator in the commutative case. The same result can be proved by extending
the symbolic approach used in \cite{mr2001h:37147} for 
the $(1+1)$-dimensional case.

\subsection{The noncommutative case}\label{Sec42}
The extension of the concept of quasi-locality to the noncommutative case
is rather complicated.
$D_x$ and $D_y$ are the only derivations for the commutative differential
ring. The extension simply enables us to apply $D_x^{-1}$ and $D_y^{-1}$
on the derivations. 
We know that the $C_f$, cf. (\ref{cmt}), are also derivations for a 
noncommutative associative algebra and we need to take them into consideration.
Based on Proposition \ref{Pro3}, we propose the following extension
using the symbolic representation.
We start with $\cA^k$, the space of polynomials in $2k$ variables,
$\xi_i$ and $\eta_i$, $i=1, \cdots, k$.
Let $\xi_i=\lambda z_i$ and $\eta_i=\mu z_i$, 
where $\lambda$, $\mu$ are constants.
We define $I(\cA^k)$ as a set of all $f\in \cA^k$ satisfying 
$f\big |_{z_1+\cdots z_k=0}=0$. 
For the elements in $g\in I(\cA^k)$, we compute $\frac{g}{\xi_1+\cdots +\xi_k}$
and $\frac{g}{\eta_1+\cdots \eta_k}$ and collect them in the set 
${\cal E}xt(\cA^k)$.
Now we define a sequence of extensions of $\cA=\cup_k \cA^k$.
Let $\cA_0^{ext}=\cA$
and $\cA_1^{ext} $ be the ring closure of the union 
$\cA_0^{ext}\cup (\cup_k {\cal E}xt(\cA^k))$. The number of variables
is a natural grade on $\cA_1^{ext}$ and again denote ${\cA_1^{ext}}^k$
the space of elements in $\cA_1^{ext}$ with $2k$ variables, 
$\xi_i$ and $\eta_i$, $i=1, \cdots, k$.
In general, we can define $\cA_l^{ext} $ is the ring closure of the union
$\cA_{l-1}^{ext}\cup (\cup_k {\cal E}xt({\cA_{l-1}^{ext}}^k))$.
Clearly, we have $\cA_{l-1}^{ext}\subset \cA_l^{ext}$.
We now define $\cA^{ext}=\lim_{l\to \infty}\cA_l^{ext}$. 

The above ring extension is bigger than we require. For example, 
the expression $\frac{\eta_2}{\xi_2}-\frac{\eta_1}{\xi_1}$ 
is not equal to zero under the substitution $\xi_1+\xi_2+\xi_3=0$
and $\eta_1+\eta_2+\eta_3=0$, that is, not satisfying
Proposition \ref{Pro3}. But it is in $I({\cA_1^{ext}}^3)$.
If we start from the symmetric polynomial, we hope to end up the
extension of commutative differential polynomial, i.e. quasi-local
polynomials we defined in section \ref{Sec2}. It is not difficult 
to notice that this extension is bigger than the concept of 
quasi-local commutative polynomials.
However, neither can we formulate the such extension using symbolic
representation nor starting from noncommutative differential
polynomials as in commutative case. It is still an open problem.

Here we compute explicitly some Lax flows for lower order scalar
Lax operators of noncommutative dependent variable, which will help
the reader to see the structures. 

\subsubsection{Noncommutative Boussinesq equation}
Let us compute the hierarchy of Lax operator $L=D_x^3+u D_x +v-D_y$.
From Proposition \ref{Pro1}, we have
\begin{eqnarray*}
&&S=X+u\frac{\xi_1^{(1)} X}{(X+\xi_1^{(1)})^3-X^3-\eta_1^{(1)}}
+v\frac{\xi_1^{(0)} }{(X+\xi_1^{(0)})^3-X^3-\eta_1^{(0)}}
\\&&
+\frac{u^2 (X+\xi_2^{(1)})X}{(X+\xi_1^{(1)}+\xi_2^{(1)})^3-X^3-\eta_1^{(1)}
-\eta_2^{(1)}}
(\frac{\xi_1^{(1)}}{(X+\xi_1^{(1)}+\xi_2^{(1)})^3-(X+\xi_2^{(1)})^3-\eta_1^{(1)}}-\frac{\xi_2^{(1)}}{(X+\xi_2^{(1)})^3-X^3-\eta_2^{(1)}})
\\&&
+\frac{uv(X+\xi_1^{(0)})}{(X+\xi_1^{(1)}+\xi_1^{(0)})^3-X^3-\eta_1^{(1)}
-\eta_1^{(0)}}
(\frac{\xi_1^{(1)}}{(X+\xi_1^{(1)}+\xi_1^{(0)})^3-(X+\xi_1^{(0)})^3-\eta_1^{(1)}}-\frac{\xi_1^{(0)}}{(X+\xi_1^{(0)})^3-X^3-\eta_1^{(0)}})
\\&&
+\frac{vuX}{(X+\xi_1^{(1)}+\xi_1^{(0)})^3-X^3-\eta_1^{(1)}
-\eta_1^{(0)}}
(\frac{\xi_1^{(0)}}{(X+\xi_1^{(1)}+\xi_1^{(0)})^3-(X+\xi_1^{(1)})^3-\eta_1^{(0)}}-\frac{\xi_1^{(1)}}{(X+\xi_1^{(1)})^3-X^3-\eta_1^{(1)}})
\\&&
+\cdots \cdots
\\&&
=X+\frac{u}{3X}+\frac{v-u\xi_1^{(1)}}{3X^2}
+(\frac{2u {\xi_1^{(1)}}^2}{9}+\frac{u \eta_1^{(1)}}{9 \xi_1^{(1)}}
-\frac{v \xi_1^{(0)}}{3}-\frac{u^2}{9})\frac{1}{X^3}
\end{eqnarray*}
\begin{eqnarray*}
&&
+(\frac{-u{\xi_1^{(1)}}^3}{9}-\frac{2 u \eta_1^{(1)}}{9}
+\frac{2 v{\xi_1^{(0)}}^2}{9}+\frac{v \eta_1^{(0)}}{9 \xi_1^{(0)}}
-\frac{uv}{9}-\frac{vu}{9}
+u^2(\frac{7 \xi_2^{(1)}}{27} +\frac{5 \xi_1^{(1)}}{27}+
\frac{\frac{\eta_1^{(1)}}{27 \xi_1^{(1)}}
-\frac{\eta_2^{(1)}}{27 \xi_2^{(1)}}}{\xi_1^{(1)}+\xi_2^{(1)}})
)\frac{1}{X^4}
+\cdots,
\end{eqnarray*}
which corresponds to 
\begin{eqnarray*}
S&=&D_x+\frac{u}{3}D_x^{-1}+(\frac{v}{3}-\frac{u_x}{3})D_x^{-2}
+(\frac{2u_{xx}}{9}+\frac{\Theta u}{9}-\frac{v_x}{3}-\frac{u^2}{9})D_x^{-3}
\\&&
+(-\frac{u_{xxx}}{9}-\frac{2u_y}{9}+\frac{2v_{xx}}{9}+\frac{\Theta v}{9}
-\frac{uv}{9}-\frac{vu}{9}+\frac{7 u u_x}{27}+\frac{5 u_x u}{27}
+\frac{ D_x^{-1}( (\Theta u) u)}{27}-\frac{ D_x^{-1}(u \Theta u)}{27})D_x^{-4}+\cdots
\end{eqnarray*}
This leads to
\begin{eqnarray*}
&&S_{\geq 0}=D_x;\quad\quad S^2_{\geq 0}=D_x^2+\frac{2}{3}u;
\quad\quad S^3_{\geq 0}=D_x^3+u D_x+v;
\\&&
S^4_{\geq 0}=D_x^4+\frac{4}{3} u D_x^2
+\frac{2}{3}(u_x+2 v) D_x +\frac{4}{9}\Theta u +\frac{2}{9}u^2
+\frac{2}{9}u_{xx}+\frac{2}{3}v_x;
\\&&
S^5_{\geq 0}=D_x^5+\frac{5}{3} u D_x^3
+\frac{5}{3}(u_x+ v) D_x^2 +(\frac{5}{9}\Theta u +\frac{5}{9}u^2
+\frac{10}{9}u_{xx}+\frac{5}{3}v_x) D_x
\\&&\quad\quad
+\frac{10}{9}v_{xx}+\frac{5}{9}(\Theta v+ uv + vu)
+\frac{5}{27}(u u_x-u_x u +D_x^{-1} ((\Theta u) u
-u \Theta u)).
\end{eqnarray*}
And we have
\begin{eqnarray*}
&&\left\{\begin{array}{l}u_{t_2}=-u_{xx}+2 v_x\\
v_{t_2}=v_{xx}+\frac{2}{3}(u_y-u_{3x}-u u_x +u v- vu)\end{array}\right.
\end{eqnarray*}
Eliminating the dependent variable $v$ from this equation and writing $t_2=t$, 
we can derive the $(2+1)$-dimensional Boussinesq equation
$$
u_{tt}=-\frac{1}{3}u_{4x}+\frac{4}{3}u_{xy}-\frac{2}{3}(u^2)_{xx}
+\frac{2}{3} D_x [u,\ D_x^{-1} u_t] .
$$
Under the scaling transformation $x\mapsto \sqrt 6 x,
\ y\mapsto \sqrt \frac{3}{2} y,\
u\mapsto 3 u,\ t\mapsto 2 t$, it can be rewritten as
$$
u_{tt}=-u_{4x}+u_{xy}-3(u^2)_{xx} + D_x [u,\ D_x^{-1} u_t] .
$$
Let $w=D_x^{-1}u$. We get $(2+1)$-dimensional noncommutative potential
Boussinesq equation
$$
w_{tt}=-w_{4x}+w_{xy}-3 (w_x^2)_{x} + [w_x,\ w_t]
$$

\subsubsection{Noncommutative KP equation}
Let us compute the hierarchy of the Lax operator $L=D_x^2+u-D_y$.
Since the order of $L$ is low, it is easier to compute 
the operator $S$ directly order by order
as in Exercise \ref{Ex4} than to apply Proposition \ref{Pro2}.
Let
$$S=D_x+a_{-1}D_x^{-1}+a_{-2}D_x^{-2}+a_{-3}D_x^{-3}\cdots .$$
Using formula (\ref{SA}), we obtain
\begin{eqnarray*}
&&a_{-1}=\frac{u}{2}\\
&&a_{-2}=-\frac{u_x}{4}+\frac{\Theta u}{4}\\
&&a_{-3}=\frac{\Theta^2 u}{8}+\frac{u_{xx}}{8}-\frac{u_y}{4}-\frac{u^2}{8}
-\frac{1}{8} D_x^{-1}C_u \Theta u
\end{eqnarray*}
This leads to
\begin{eqnarray*}
&&S^3_{\geq 0}=D_x^3+\frac{3 u}{2} D_x+\frac{3}{4}
 ( u_x+\Theta u)\\
&&S^4_{\geq 0}=D_x^4+2 u D_x^2+( 2 u_x +\Theta u) D_x
 +\frac{u_y}{2}+u^2+u_{xx}+\frac{\Theta^2 u}{2}-\frac{1}{2}D_x^{-1}C_u \Theta u
\\&&
=2 D_x^2 D_y -D_y^2+2 uD_y+ \Theta u D_x
 +\frac{3 u_y}{2}+\frac{\Theta^2 u}{2}-\frac{1}{2}D_x^{-1}C_u \Theta u
+L^2 .
\end{eqnarray*}
And we have
\begin{eqnarray*}
u_{t_3}=\frac{1}{4} (u_{xxx} +3 u u_x +3 u_x u +3 \Theta u_y -3 C_u \Theta u ),
\end{eqnarray*}
which is the noncommutative KP equation and can be obtained from the recursion
operator of KP, cf. \cite{mr93c:58085},
and 
\begin{eqnarray*}
&&u_{t_4}=\frac{1}{2} (u_{xxy}+\Theta^2 u_y+2 u u_y +2 u_y u+u_x \Theta u
+(\Theta u) u_x
 +C_u D_x^{-1} C_u \Theta u
-\Theta C_u \Theta u-C_u \Theta^2 u) ,
\end{eqnarray*}
which appeared in \cite{HT03}, cf. formula (2.14), with a mistake:
the last two terms are missing.

\subsubsection{Noncommutative mKP equation}
Let us compute the hierarchy of the Lax operator $L=D_x^2+u D_x +v+D_x^{-1} w-D_y$.
Let
$$S=D_x+a_0+a_{-1}D_x^{-1}+\cdots .$$
Using formula (\ref{SA}), we obtain
\begin{eqnarray*}
&&a_0=\frac{u}{2}\\
&&a_{-1}=-\frac{u_x}{4}-\frac{u^2}{8}+\frac{v}{2}+(D_x+\frac{C_u}{2})^{-1} 
\frac{u_y}{4}
\end{eqnarray*}
This leads to
\begin{eqnarray*}
&&S_{\geq 1}=D_x;\quad\quad S^2_{\geq 1}=D_x^2+uD_x;
\quad\quad S^3_{\geq 1}=D_x^3+\frac{3 u}{2} D_x^2+\frac{3}{8}
 (2 u_x+u^2+4 v+2 (D_x+\frac{C_u}{2})^{-1} u_y)D_x.
\end{eqnarray*}
And we have
\begin{eqnarray*}
&&\left\{\begin{array}{l}u_{t_2}=u_y+2 v_x+uv-vu
\\
v_{t_2}=v_{x}+2 w_x+u v_x +uw-wu
\\
w_{t_2}=-w_{xx}+(w u)_{x}\end{array}\right.
\end{eqnarray*}
\begin{eqnarray*}
&&\left\{\begin{array}{l}u_{t_3}=\frac{1}{4} u_{xxx}+\frac{3}{8}C_u u_{xx}
-\frac{3}{8} u u_x u +\frac{3}{4} u_{xy}+\frac{3}{8} (u^2)_y
+\frac{3}{2}v_{xx}+\frac{3}{2}v_y
+\frac{3}{2}u v_x +\frac{3}{2} v u_x-\frac{3}{8}C_v u^2
\\ \quad\quad
-\frac{3}{4}C_v u_x
+3 w_x+\frac{3}{2}C_u w
+\frac{3}{4} (D_y -D_x^2-u D_x+R_{u_x}-C_v)(D_x+\frac{C_u}{2})^{-1} u_y
\\
v_{t_3}=v_{xxx}+\frac{3}{2}u v_{xx}+\frac{3}{4}u_x v_x+\frac{3}{8}u^2 v_1
+\frac{3}{2}v v_1+\frac{3}{2}u w_x+\frac{3}{2} w_x u+\frac{3}{4} w u_x
+\frac{3}{4} u_x w-\frac{3}{8} C_w u^2
\\ \quad\quad
-\frac{3}{2} C_w v +\frac{3}{4} (R_{v_x}-C_w) (D_x+\frac{C_u}{2})^{-1} u_y\\
w_{t_3}=w_{xxx}-(w a_0)_{xx}+(w a_{-1})_x\end{array}\right.
\end{eqnarray*}
The reduction $v=w=0$ leads to the noncommutative mKP equation
\begin{eqnarray}
u_{t_3}=\frac{1}{4} u_{xxx}+\frac{3}{8}C_u u_{xx}
-\frac{3}{8} u u_x u +\frac{3}{4} u_{xy}+\frac{3}{8} (u^2)_y
+\frac{3}{4} (D_y -D_x^2-L_u D_x+R_{u_x})(D_x+\frac{C_u}{2})^{-1} u_y .
\label{nmkp}
\end{eqnarray}
If we introduce new variable $p$ satisfying  $u_y=p_x+\frac{1}{2}C_u p $,
The system for $u$ and $p$ is equivalent to the matrix mKP in 
\cite{Adler}.

If $u$ does not depend on $y$, i.e. $u_y=0$, this gives us the
noncommutative  mKdV equation, \cite{mr99c:58077, mr1781148}
$$
u_{t_3}=\frac{1}{4} u_{xxx}+\frac{3}{8}C_u u_{xx} -\frac{3}{8} u u_x u .
$$
If $u$ takes its value over $\bbbc$ (commutative), this gives us
the well known mKP equation
\begin{eqnarray}
u_{t_3}=\frac{1}{4} u_{xxx} -\frac{3}{8} u^2 u_x 
+\frac{3}{4} \Theta u_{y}
+\frac{3}{4} u_x \Theta u .
\label{mkp}
\end{eqnarray}
We know Miura transformation $V=u_x-\frac{u^2}{2}+\Theta u$ transforms
the mKP equation (\ref{mkp}) into the KP equation 
$V_{t_3}=\frac{1}{4} V_{xxx}+\frac{3}{4} V V_x+\frac{3}{4} \Theta V_y$.
However, we do not know a Miura transformation for noncommutative mKP
(\ref{nmkp}).

\subsubsection{Nonsymmetric Novikov-Veselov equation}
Consider the Lax operator $L=u + D_x^{-1} v  -D_y$ (case {\bf c})
\footnote{In the commutative case, this worked out in paper \cite{BS01} 
by both central extension approach and the operand approach.}.
Let $S=D_x+a_0+a_{-1}D_x^{-1}+\cdots$.
Using formula (\ref{S}), we have 
\begin{eqnarray*}
&& a_0=(C_u-D_y)^{-1} u_x;\\ 
&&a_{-1}=(C_u-D_y)^{-1} (v_x+C_v a_0),
\end{eqnarray*}
leading to ${S^1}_{\geq 1}=D_x$,\quad  ${S^2}_{\geq 1}=D_x^2+2 a_0 D_x$,\quad
${S^3}_{\geq 1}=D_x^3 +3a_0 D_x^2+3(a_{0x}+ a_{-1}+a_{0}^2)D_x$. 
From the Lax equation, we obtain
\begin{eqnarray*}
&&\left\{\begin{array}{l}
u_{t_2}=u_{xx}+2 v_x+2 a_0 u_x+2 a_0 v-2 v a_0\\
v_{t_2}=-v_{xx}+2(v a_0)_x
\end{array}\right.
\\&&
\left\{\begin{array}{l}
u_{t_3}= u_{xxx}+3 (a_0 u_{x})_{x}+3 a_{-1} u_x +3 a_0^2 u_x
+3a_0 v_x+3(v a_0)_x-3C_v(a_{0x}+ a_{-1}+a_{0}^2)\\
v_{t_3}=v_{xxx}-3(v_x a_0)_x+3 (v a_{-1})_x+3 (v a_0^2)_x
\end{array}\right.
\end{eqnarray*}
The reduction $v=0$ leads to 
$$u_{t_2}=u_{xx}-2 \{(D_y-C_u)^{-1} u_x\} u_x .$$
Let $w=(D_y-C_u)^{-1} u_x$. This equation transforms into
the noncommutative Burgers equation $w_{t_2}=w_{xx}-2 w w_x$,
which is linearisable, that is, we obtain $p_t=p_{xx}$ by $pw=-p_x$.

There is no reduction $u=0$ in noncommutative case.
In the commutative case, the reduction $u=0$ leads to 
the nonsymmetric Novikov--Veselov equation \cite{NV86}
$$v_{t_3}=v_{xxx}-3(v \Theta^{-1} v)_x .$$
\section{Conclusion}
We have demonstrated the power of the symbolic representation in the study of
$(2+1)$-dimensional integrable equations. 
It enables us to produce the hierarchies of $(2+1)$-dimensional 
integrable equations from the  
scalar Lax operators and to study their properties globally. 
We proved the conjecture of Mikhailov and Yamilov on the ring extension for 
$(2+1)$-dimensional commutative integrable equations derived
from scalar Lax operators. The method can easily adapt to the case
of $(2+1)$-dimensional differential-difference integrable equation with scalar Lax operators \cite{BS01}.

Noncommutative integrable equations can be considered as quantised analogs
of classical integrable equations, which is very important in string theory.
Right after finishing the paper, we noticed a few papers on constructing
noncommutative integrable equations in the framework of the Sato theory
motivated by noncommutative gauge theories,
cf. \cite{HT03,HM05} and references therein.
However, the authors only used an ansatz to derive the equations
from some scalar Lax operators in case {\bf a} instead of this
systematical manner.
There are few papers on noncommutative $(2+1)$-dimensional integrable equations
derived from the scalar Lax operators in cases {\bf b} and {\bf c}
such as noncommutative mKP, where
the new type of nonlocal terms such as 
$(D_x+  \frac{C_u}{2})^{-1} u_y$ and $(D_y-  C_u)^{-1} u_x$ appear.
To study these equations further such as constructing their Hamiltonian 
operators and soliton solutions may provide a deeper understanding
of noncommutative geometry.

\section*{Acknowledgments}
The author would like to thank  
A.~Hone, 
A.V. Mikhailov, V.S. Novikov and J.A.~Sanders for useful discussions.

\section*{Appendix: Proof of Proposition \ref{Pro3}}
{\bf Proof.} We need to show (cf. (\ref{Bf}) for all $l\geq 2$ that
\begin{eqnarray}
b_{l-1}(-\sum_{s=2}^l\xi_{j_s}^{(i_s)},-\sum_{s=2}^l\eta_{j_s}^{(i_s)},
\xi_{j_2}^{(i_2)},\eta_{j_2}^{(i_2)},\cdots,\xi_{j_{l-1}}^{(i_{l-1})},
\eta_{j_{l-1}}^{(i_{l-1})},X+\xi_{j_{l}}^{(i_{l})})
=b_{l-1}(\xi_{j_2}^{(i_2)},\eta_{j_2}^{(i_2)},\cdots,\xi_{j_l}^{(i_{l})},
\eta_{j_l}^{(i_{l})},X).
\label{Cond}
\end{eqnarray}
We prove it by induction. It is true for $l=2$ since
\begin{eqnarray*}
&&
b_1(-\xi_{j_2}^{(i_2)},-\eta_{j_2}^{(i_2)},X+\xi_{j_2}^{(i_2)})
=\frac{-\xi_{j_2}^{(i_2)}}{N_1(-\xi_{j_2}^{(i_2)},-\eta_{j_2}^{(i_2)},X+\xi_{j_2}^{(i_2)})}
=\frac{\xi_{j_2}^{(i_2)}}{N_1(\xi_{j_2}^{(i_2)},\eta_{j_2}^{(i_2)},X)}
=b_1(\xi_{j_2}^{(i_2)},\eta_{j_2}^{(i_2)},X).
\end{eqnarray*}
Assume that formula (\ref{Cond}) is valid for $l-1$.
Then its the left-hand side equals
\begin{eqnarray*}
&&
\frac{b_{l-2}(-\sum_{s=2}^l\xi_{j_s}^{(i_s)},-\sum_{s=2}^l\eta_{j_s}^{(i_s)},
\xi_{j_2}^{(i_2)},\eta_{j_2}^{(i_2)}\cdots,\xi_{j_{l-2}}^{(i_{l-2})},
\eta_{j_{l-2}}^{(i_{l-2})}, X+\xi_{j_{l-1}}^{(i_{l-1})}+\xi_{j_{l}}^{(i_l)})
}
{N_{l-1}(-\sum_{s=2}^l\xi_{j_s}^{(i_s)},-\sum_{s=2}^l\eta_{j_s}^{(i_s)},
\xi_{j_2}^{(i_2)},\eta_{j_2}^{(i_2)},\cdots,\xi_{j_{l-1}}^{(i_{l-1})},
\eta_{j_{l-1}}^{(i_{l-1})},X+\xi_{j_{l}}^{(i_{l})})}
\\&&
-\frac{b_{l-2}(\xi_{j_2}^{(i_2)},\eta_{j_2}^{(i_2)},\cdots, \xi_{j_{l-1}}^{(i_{l-1})},\eta_{j_{l-1}}^{(i_{l-1})},X+\xi_{j_{l}}^{(i_{l})})}
{N_{l-1}(-\sum_{s=2}^l\xi_{j_s}^{(i_s)},-\sum_{s=2}^l\eta_{j_s}^{(i_s)},
\xi_{j_2}^{(i_2)},\eta_{j_2}^{(i_2)},\cdots,\xi_{j_{l-1}}^{(i_{l-1})},
\eta_{j_{l-1}}^{(i_{l-1})},X+\xi_{j_{l}}^{(i_{l})})}
\\&=&
-\frac{b_{l-2}(\xi_{j_2}^{(i_2)},\eta_{j_2}^{(i_2)}\cdots,\xi_{j_{l-2}}^{(i_{l-2})},
\eta_{j_{l-2}}^{(i_{l-2})},\xi_{j_{l-1}}^{(i_{l-1})}+\xi_{j_{l}}^{(i_l)},
\eta_{j_{l-1}}^{(i_{l-1})}+\eta_{j_{l}}^{(i_l)}, X)
}
{N_{1}(\xi_{j_l}^{(i_{l})}, \eta_{j_l}^{(i_{l})},X)}
\\&&
+\frac{
b_{l-2}(\xi_{j_2}^{(i_2)},\eta_{j_2}^{(i_2)},\cdots, \xi_{j_{l-1}}^{(i_{l-1})},\eta_{j_{l-1}}^{(i_{l-1})},X+\xi_{j_{l}}^{(i_{l})})}
{N_{1}(\xi_{j_l}^{(i_{l})}, \eta_{j_l}^{(i_{l})},X)}
\\
&=&
\frac{b_{l-3}(\xi_{j_2}^{(i_2)},\eta_{j_2}^{(i_2)}\cdots,\xi_{j_{l-2}}^{(i_{l-2})},
\eta_{j_{l-2}}^{(i_{l-2})}, X+\xi_{j_{l-1}}^{(i_{l-1})}+\xi_{j_{l}}^{(i_{l})})
-b_{l-3}(\xi_{j_3}^{(i_3)},\eta_{j_3}^{(i_3)},\cdots, \xi_{j_{l-1}}^{(i_{l-1})},\eta_{j_{l-1}}^{(i_{l-1})},X+\xi_{j_{l}}^{(i_{l})})}
{N_{l-2}(\xi_{j_2}^{(i_2)},\eta_{j_2}^{(i_2)},\cdots,\xi_{j_{l-1}}^{(i_{l-1})},\eta_{j_{l-1}}^{(i_{l-1})},
X+\xi_{j_{l}}^{(i_{l})})
N_1(\xi_{j_l}^{(i_{l})},\eta_{j_l}^{(i_{l})},X)}
\\&&
-\frac{b_{l-3}(\xi_{j_2}^{(i_2)},\eta_{j_2}^{(i_2)}\cdots,\xi_{j_{l-2}}^{(i_{l-2})},
\eta_{j_{l-2}}^{(i_{l-2})}, X+\xi_{j_{l-1}}^{(i_{l-1})}+\xi_{j_{l}}^{(i_l)})}
{N_{l-2}(\xi_{j_2}^{(i_2)},\eta_{j_2}^{(i_2)},\cdots,\xi_{j_{l-1}}^{(i_{l-1})}+\xi_{j_{l}}^{(i_l)},
\eta_{j_{l-1}}^{(i_{l-1})}+\eta_{j_{l}}^{(i_l)},X)
N_1(\xi_{j_l}^{(i_{l})},\eta_{j_l}^{(i_{l})},X)}
\\&&
+\frac{b_{l-3}(\xi_{j_3}^{(i_3)},\eta_{j_3}^{(i_3)},\cdots,
\xi_{j_{l-1}}^{(i_{l-1})}+\xi_{j_{l}}^{(i_l)},\eta_{j_{l-1}}^{(i_{l-1})}+\eta_{j_{l}}^{(i_l)},X)}
{N_{l-2}(\xi_{j_2}^{(i_2)},\eta_{j_2}^{(i_2)},\cdots,\xi_{j_{l-1}}^{(i_{l-1})}+\xi_{j_{l}}^{(i_l)},
\eta_{j_{l-1}}^{(i_{l-1})}+\eta_{j_{l}}^{(i_l)},X)
N_1(\xi_{j_l}^{(i_{l})},\eta_{j_l}^{(i_{l})},X)};
\end{eqnarray*}
Meanwhile, the right side gives us
\begin{eqnarray*}
&&\frac{b_{l-2}(\xi_{j_2}^{(i_2)},\eta_{j_2}^{(i_2)}\cdots,\xi_{j_{l-1}}^{(i_{l-1})},
\eta_{j_{l-1}}^{(i_{l-1})}, X+\xi_{j_{l}}^{(i_l)})
-b_{l-2}(\xi_{j_3}^{(i_3)},\eta_{j_3}^{(i_3)},\cdots, \xi_{j_l}^{(i_l)},\eta_{j_l}^{(i_l)},X)}
{N_{l-1}(\xi_{j_2}^{(i_2)},\eta_{j_2}^{(i_2)},\cdots,\xi_{j_l}^{(i_l)},\eta_{j_l}^{(i_l)},X)}
\\&=&
\frac{b_{l-3}(\xi_{j_2}^{(i_2)},\eta_{j_2}^{(i_2)},\cdots,
\xi_{j_{l-2}}^{(i_{l-2})},\eta_{j_{l-2}}^{(i_{l-2})},X+\xi_{j_{l-1}}^{(i_{l-1})}+\xi_{j_{l}}^{(i_l)})}
{N_1(\xi_{j_{l-1}}^{(i_{l-1})},\eta_{j_{l-1}}^{(i_{l-1})},X+\xi_{j_{l}}^{(i_l)})N_{l-1}(\xi_{j_2}^{(i_2)},\eta_{j_2}^{(i_2)},\cdots,\xi_{j_l}^{(i_l)},\eta_{j_l}^{(i_l)},X)}
\\&&
-\frac{b_{l-3}(\xi_{j_2}^{(i_2)},\eta_{j_2}^{(i_2)},\cdots,
\xi_{j_{l-2}}^{(i_{l-2})}+\xi_{j_{l-1}}^{(i_{l-1})},\eta_{j_{l-2}}^{(i_{l-2})}+\eta_{j_{l-1}}^{(i_{l-1})},X+\xi_{j_{l}}^{(i_l)})}
{N_1(\xi_{j_{l-1}}^{(i_{l-1})},\eta_{j_{l-1}}^{(i_{l-1})},X+\xi_{j_{l}}^{(i_l)})N_{l-1}(\xi_{j_2}^{(i_2)},\eta_{j_2}^{(i_2)},\cdots,\xi_{j_l}^{(i_l)},\eta_{j_l}^{(i_l)},X)}
\\&-&
\frac{b_{l-3}(\xi_{j_3}^{(i_3)},\eta_{j_3}^{(i_3)},\cdots,
\xi_{j_{l-1}}^{(i_{l-1})},\eta_{j_{l-1}}^{(i_{l-1})},X+\xi_{j_{l}}^{(i_{l})})
-b_{l-3}(\xi_{j_3}^{(i_3)},\eta_{j_3}^{(i_3)},\cdots,
\xi_{j_{l-1}}^{(i_{l-1})}+\xi_{j_{l}}^{(i_{l})},\eta_{j_{l-1}}^{(i_{l-1})}+\eta_{j_{l}}^{(i_{l})},X)}
{N_1(\xi_{j_l}^{(i_{l})},\eta_{j_l}^{(i_{l})},X)
N_{l-1}(\xi_{j_2}^{(i_2)},\eta_{j_2}^{(i_2)},\cdots,\xi_{j_l}^{(i_l)},\eta_{j_l}^{(i_l)},X)}.
\end{eqnarray*}
The difference between the above two expression vanishes due to the
induction assumption, which implies that
\begin{eqnarray*}
&&
\frac{
N_{l-3}(\xi_{j_2}^{(i_2)},\eta_{j_2}^{(i_2)},\cdots,\xi_{j_{l-2}}^{(i_{l-2})},
\eta_{j_{l-2}}^{(i_{l-2})}, X+\xi_{j_{l-1}}^{(i_{l-1})})
b_{l-3}(\xi_{j_2}^{(i_2)},\eta_{j_2}^{(i_2)}\cdots,\xi_{j_{l-2}}^{(i_{l-2})},
\eta_{j_{l-2}}^{(i_{l-2})}, X+\xi_{j_{l-1}}^{(i_{l-1})})
}
{N_{l-2}(\xi_{j_2}^{(i_2)},\eta_{j_2}^{(i_2)},\cdots,\xi_{j_{l-1}}^{(i_{l-1})},
\eta_{j_{l-1}}^{(i_{l-1})}, X)
N_1(\xi_{j_{l-1}}^{(i_{l-1})},\eta_{j_{l-1}}^{(i_{l-1})},X)
}
\\&=&
\frac{b_{l-3}(\xi_{j_2}^{(i_2)},\eta_{j_2}^{(i_2)},\cdots,
\xi_{j_{l-2}}^{(i_{l-2})}+\xi_{j_{l-1}}^{(i_{l-1})},\eta_{j_{l-2}}^{(i_{l-2})}+\eta_{j_{l-1}}^{(i_{l-1})},X)}
{N_1(\xi_{j_{l-1}}^{(i_{l-1})},\eta_{j_{l-1}}^{(i_{l-1})},X)
}
-\frac{b_{l-3}(\xi_{j_3}^{(i_3)},\eta_{j_3}^{(i_3)},\cdots,
\xi_{j_{l-1}}^{(i_{l-1})},\eta_{j_{l-1}}^{(i_{l-1})},X)
}
{N_{l-2}(\xi_{j_2}^{(i_2)},\eta_{j_2}^{(i_2)},\cdots,\xi_{j_{l-1}}^{(i_{l-1})},
\eta_{j_{l-1}}^{(i_{l-1})}, X)
}.
\end{eqnarray*}
By now we proved our proposition.
$\diamond$

\end{document}